# Teaching Physics Using Virtual Reality

C. Savage[*], D. McGrath[†], T. McIntyre[¶], M. Wegener[¶], M. Williamson[*]

[*]*Centre for Learning and Teaching in the Physical Sciences, The Australian National University, Canberra, ACT 0200, Australia*
[†]*The Teaching and Educational Development Institute, The University of Queensland, Brisbane 4072, Australia*
[¶]*School of Physics, The University of Queensland, Brisbane 4072, Australia.*

**Abstract.** We present an investigation of game-like simulations for physics teaching. We report on the effectiveness of the interactive simulation "Real Time Relativity" for learning special relativity. We argue that the simulation not only enhances traditional learning, but also enables new types of learning that challenge the traditional curriculum. The lessons drawn from this work are being applied to the development of a simulation for enhancing the learning of quantum mechanics.

## INTRODUCTION

Computer simulations are routinely used to inform critical decisions such as those concerning the economy and the environment. They are also used in physics education [1] and in relativity education in particular [2]. In this paper we explore ways to make further use of computer simulations in physics education.

Computer games occupy a central place in society [3]. Many people, especially young people, have a great deal of experience learning the rules of game-worlds. In some cases, such as the first-person-shooter genre, this includes learning the physics of the simulated world through a process of experimentation. One of the questions that motivated our work is: can physics educators take advantage of this experience to help students learn physics?

We present a case study of the use of a simulation in special relativity education, and a preliminary study for quantum mechanics education. These two areas are at the abstract end of physics, in that students lack direct experience of them. Indeed they both challenge deeply ingrained notions about how the world works.

They are also often taught with an emphasis on their mathematical formulation rather than on their experimental basis. Indeed special relativity is sometimes taught precisely because it provides a beautiful example of an axiomatic theory in physics.

We argue that computer simulations can complement traditional approaches to teaching by providing simulated experiences of unfamiliar physics. We also suggest that such simulations emphasise different physics to that in typical lecture courses and hence may stimulate re-evaluation of the curriculum.

## SPECIAL RELATIVITY

Learning special relativity requires students to reconcile fundamentally new notions of space and time with their everyday conceptions. This is often unsuccessful, even for advanced students [4].

### The Simulation

We have developed a three-dimensional, first person, interactive, game-like simulation of relativistic physics called "Real Time Relativity" (RTR) [5,6]. It provides a number of scenarios ranging from realistic planets to fantasy cityscapes: Fig. 1.

The simulation encompasses relativistic kinematics and optics, including time dilation and the relativity of simultaneity. The relativistic physics seen in Fig. 1 is aberration, which was discussed in Einstein's 1905 relativity paper [7]. The Doppler and headlight effects have been suppressed for clarity. The relevant physics is explained in detail in references [5] and [8].

With the exception of planets, familiar objects are too small for relativistic visual effects to be noticeable. Hence the objects portrayed in some of our scenarios, such as buildings, are unrealistically large. However ensuring an explicit and consistent scale within each scenario has avoided any significant source of confusion for students.

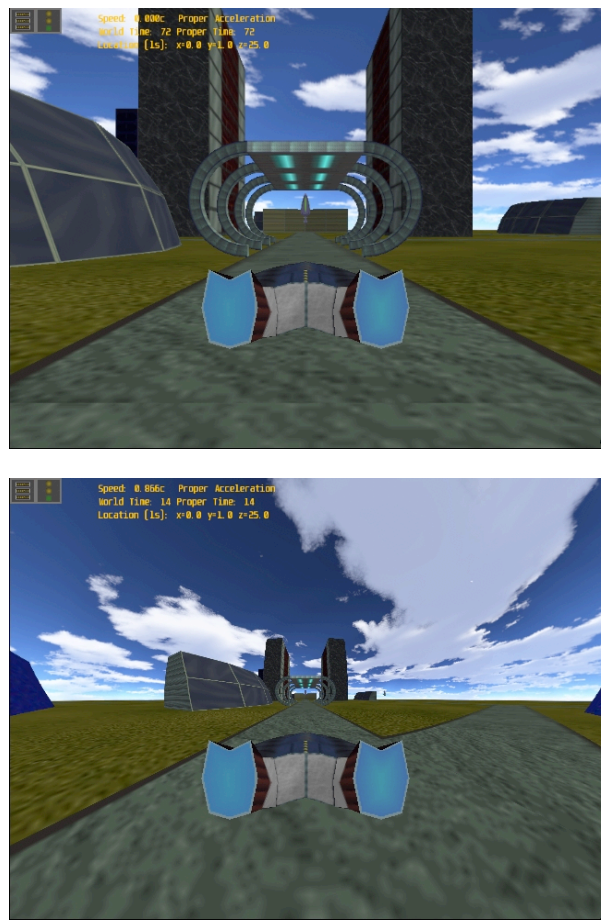

**FIGURE 1.** Screenshots from the cityscape scenario of RTR. Top frame: observer riding the rocket is at rest relative to the cityscape. Bottom frame: observer riding the rocket is travelling at $0.87c$ relative to the cityscape. Both images correspond to the same event, the only difference being the speed of the observer at that event.

## Using Real Time Relativity in University Courses

The Real Time Relativity simulation has been used in a number of first year physics courses in Australia. At the Australian National University and at The University of Queensland it has been the basis of a three-hour computer lab that was part of otherwise conventional relativity instruction.

Over a number of years various styles of laboratory were tried and evaluated. This process converged on a balance between directed activities and open-ended exploration, with a strong conceptual emphasis [6]. Feedback from students' experience of these labs was also used to develop the simulation's interface and capabilities.

## Evaluation of Effectiveness

We have conducted extensive evaluations of the educational effectiveness of the Real Time Relativity labs [9]. This has included pre- and post-lab testing of the understanding and attitudes of hundreds of students, as well as ethnographic observation of student behaviour during the labs.

We surveyed a class at The Australian National University in 2008. At the time of the survey 19 students had completed the RTR lab and 39 had not. Comparing these two groups, 77% of those who had not done the lab agreed with the statement that "special relativity is more abstract than most topics in physics", while only 55% of those who had done the lab agreed. Hence using RTR makes relativity less abstract for students

For a sample of 200 students from The Australian National University and The University of Queensland who had completed the RTR lab we found that 77% found the lab to be interesting and that 70% would like to use more simulations in their studies. This demonstrates that students believe the simulation supports their learning.

Concept tests were given to students before and after the RTR lab. The responses showed statistically significant improvements in questions relating to time dilation and to the relativity of simultaneity, which are conceptually challenging aspects of relativity.

We also found that amongst students who had completed the same lecture course those who had completed the RTR lab performed better on exam questions than those who had not. A class of 182 was randomly assigned to a group of 132 students who did the lab or to a group of 50 who did not. The written exam for the course included questions on relativity that were set and marked by people unassociated with the RTR project. The mean mark out of ten for these questions was 5.4 for the RTR group and 4.6 for the non-RTR group. This is a small but statistically significant difference with a probability of less than 5% of occurring by chance.

Students engaged with highly visual phenomena, such as the headlight effect, that are very obvious in RTR but are not often included in first-year curricula. In response to an open-ended survey question about what they found most interesting about using the simulation, students most commonly referred to optical effects.

# QUANTUM MECHANICS

We are applying what we have learnt from developing the Real Time Relativity teaching package to developing a simulation of quantum dynamics. The objective is to give students an experience of a simulated world in which quantum mechanics is dominant.

Quantum dynamics is an increasingly important part of quantum mechanics due to the rise of quantum information and computing, quantum optics, and the study of ultra-cold quantum gases, such as Bose-Einstein condensates. However it is a relatively neglected part of many introductory quantum mechanics courses and texts. These tend to emphasise time-independent quantum mechanics, with perhaps some time-dependent perturbation theory. Quantum measurement is part of quantum dynamics, and is also often neglected.

## The Simulation

Designing any simulation requires deciding what will be visualized. For quantum mechanics we have chosen to visualise the single particle wavefunction. Despite disagreement in the field about its ontological status, that is about whether it should be regarded as part of reality [10], it is a fundamental quantum mechanical concept. If it is regarded as a mathematical abstraction then we are visualizing mathematics. This is quite different to Real Time Relativity, which visualized what would be seen by the eye, or by a camera. Figure 2 shows the wavefunction as visualized by our prototype software, called "QSim".

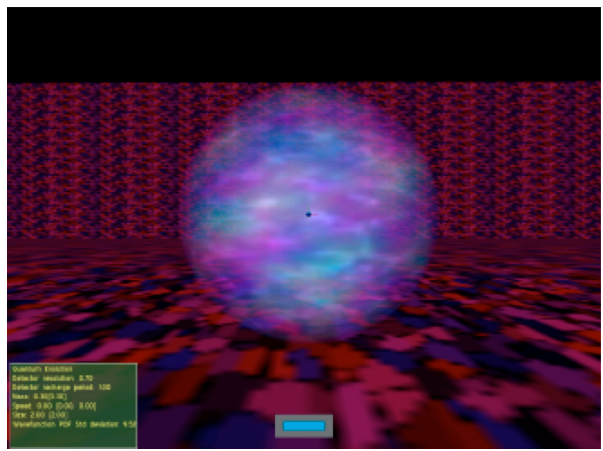

**FIGURE 2.** Screenshot from QSim. The large spherical object at centre represents an isotropic Gaussian single particle wavefunction.

The simulation evolves the wavefunction in time according to the Schrödinger equation. It also implements discrete position measurements with variable resolution [10,11]. This makes possible a number of quantum measurement activities. For example, position measurements may be repeated automatically, with a specified frequency and number of measurements, and the results plotted in three-dimensional space. This allows for an investigation of the quantum Zeno effect [10,12] at the introductory level, even though it is often regarded as an advanced topic. This echoes the relativity simulation, which allows the traditionally advanced topic of relativistic optics to be studied at the introductory level.

## Evaluation of Effectiveness

Our development process for educational simulations is based on iterative cycles of software development and evaluation with students. We report on the outcome of the first iteration of this cycle.

Key objectives of the first evaluation were to determine if students correctly interpreted the visualisation and if students felt that it improved their understanding of quantum mechanics.

Six third year physics students from The Australian National University participated in the evaluation, which consisted of a one hour structured computer laboratory. There were three activities: free evolution of the wave function, position measurement, and the quantum Zeno effect. Students were pre- and post-tested on the confidence with which they understood the relevant physics. In open-ended questions they were asked to briefly summarize in writing what they learned in each part of the lab.

On Likert scale questions, all six students rated QSim as either "enormously useful" or "quite useful" as a learning tool. When asked whether QSim improved their confidence in their understanding of specific concepts such as the Heisenberg uncertainty principle most students selected either "a little" or "significantly", with occasional selections of "a lot" and of "not at all". The overall impression from the surveys was that students felt they had learnt things, but that they weren't quite sure what. This is encouraging, as these students had already done two courses in quantum mechanics, while the target students are introductory level.

One response to an open-ended request for comments on QSim was:

> "Great idea as it enables the student to clearly visualize something which is so different from classical mechanics. Demonstrates relationships well by allowing you to change variables. Wavefunction looks cool …"

Another student when asked to comment on how their understanding of the wavefunction had changed wrote:

> "…Being given a visual representation helps build a mental picture of what is happening in quantum mechanics …"

The importance of visual models for some students is a theme we also encountered in our evaluations of Real Time Relativity: some students describe themselves as "visual learners".

## CONCLUSION

We have presented evidence that three-dimensional interactive simulations can enhance students' understanding of abstract areas of physics.

It is well known that most problems in physics can only be solved computationally. Nevertheless, many physics curricula have their origins in a time before computers were commonly available. We have argued that using computers as an integral part of the educational process opens up new possibilities for teaching at the introductory level. In particular, Real Time Relativity makes relativistic optics an accessible part of introductory relativity, and QSim makes quantum dynamics and quantum measurement accessible at the introductory level.

## ACKNOWLEDGMENTS

Support for this study has been provided by The Australian Learning and Teaching Council, an initiative of the Australian Government Department of Education, Science and Training. The views expressed in this presentation do not necessarily reflect the views of The Australian Learning and Teaching Council.

## REFERENCES

1. C. Wieman and K. Perkins, *Nat. Phys.* **2**, 290 (2006).
2. E. F. Taylor, *Am. J. Phys.* **57**, 508 (1989).
3. P. M. Greenfield, *Science* **323**, 69 (2009).
4. R. E. Scherr, P. S. Shaffer, and S. Vokos, *Am. J. Phys.* **69**, S24 (2004).
5. C. M. Savage, A. Searle, and L. McCalman, *Am. J. Phys.* **75**, 791-2504 (2007).
6. Real Time Relativity web site: http://realtimerelativity.org
7. A. Einstein, Ann. Phys. **17**, 891 (1905). Reprinted in English translation in J. Stachel, *Einstein's Miraculous Year*, Princeton University Press, Princeton, 1998.
8. Through Einstein's Eyes web site: http://www.anu.edu.au/Physics/Savage/TEE
9. D. McGrath, C. Savage, M. Williamson, M. Wegener and T. McIntyre, "Teaching Special Relativity using Virtual Reality", in *Proceedings of the UniServe Science Symposium on Visualisation and Concept Development*, http://science.uniserve.edu.au/pubs/procs/2008/index.html
10. Y. Aharonov and D. Rohrlich, *Quantum Paradoxes*, Wiley-VCH, Weinheim, 2005.
11. T. Konrad, A. Rothe, F. Petruccione, and L. Diosi, arXiv:0902.2249 (2009).
12. B. Misra and E. C. G. Sudarshan, *J. Math. Phys.* **18**, 756 (1977).